\documentclass[3p, 11pt]{elsarticle}
\pdfoutput=1
\usepackage{graphicx,amscd,amsmath,amssymb,verbatim}
\usepackage{amsfonts,epsfig}
\usepackage{mathptmx}
\usepackage{setspace}
\usepackage{epsfig}
\usepackage{url}
\usepackage{multirow}
\usepackage{subfigure}
\usepackage{color}

\begin{document}

\journal{Journal of Statistical Mechanics}

\begin{frontmatter}

\title{Numerical study of condensation in a Fermi-like model of
conterflowing particles via Gini coefficient }

\author{Eduardo V. Stock, Roberto da Silva, Carlo R. da Cunha} 

\address{Institute of Physics, Federal University of Rio Grande do Sul,
Av. Bento Gon\c{c}alves, 9500, Porto Alegre, 91501-970, RS, Brazil.}


\begin{abstract}
	The collective motion of self-driven particles shows interesting novel phenomena such as swarming and the emergence of patterns. We have recently proposed a model for counterflowing particles that captures this idea and exhibits clogging transitions. This model is based on a generalization of the Fermi-Dirac statistics wherein the maximal occupation of a cell is used. Here we present a detailed study comparing synchronous and asynchronous stochastic dynamics within this model. We show that an asynchronous updating scheme supports the mobile-clogging transition and eliminates some mobility anomalies that are present in synchronous Monte Carlo simulations. Moreover, we show that this transition is dependent upon its initial conditions. Although the Gini coefficient was originally used to model wealth inequalities, we show that it is also efficient for studying the mobile-clogging transition. Finally, we compare our stochastic simulation with direct numerical integration of partial differential equations used to describe this model.
\end{abstract}
\end{frontmatter}

\tableofcontents

\setlength{\baselineskip}{0.7cm}


\section{Introduction}

\label{Section:Introduction}

The motion of counterflowing streams of particles \cite%
{Schmittmann1992,Helbing2000,Marroquin2014} is a current important topic in
the physics of complex systems. Examples range from the motion of oppositely
charged colloidal particles\cite{Vissers2011,VissersPRL2011} to the flow of
pedestrians\cite{Helbing2000,Marroquin2014,Pinho2016}. The formation of
patterns is a general property of such systems and different models have
been proposed to explain the emergence of structures such as lanes, clogging
and jamming. \cite{Marroquin2014,Stock2017,Wei2015}.

Three dimensional flows often can be studied by one dimensional transport
models \cite{Majundar2005,Ewans2000} or decomposed in separate flows through
narrow pipes or independent cells. For instance, we have proposed a model
for the flow or counterflow of particles that is governed by a two-species
partial differential equation system (TSPDES):\cite%
{rdasilva2015,Stock2017,rdasilva2019}

\begin{equation}
\begin{array}{ccc}
\frac{\partial \rho _{A}}{\partial t} & = & K_{A}\frac{\partial }{\partial x}%
\left[ f_{A}(\rho _{A},\rho _{B})\rho _{A}\right] +D_{A}\frac{\partial ^{2}}{%
\partial x^{2}}\rho _{A} \\ 
&  &  \\ 
\frac{\partial \rho _{B}}{\partial t} & = & K_{B}\frac{\partial }{\partial x}%
\left[ f_{B}(\rho _{A},\rho _{B})\rho _{B}\right] +D_{B}\frac{\partial ^{2}}{%
\partial x^{2}}\rho _{B}%
\end{array}
\label{eq:general_equation}
\end{equation}%
where $\rho _{A,B}$ is the mass concentration, $K_{A,B}$ is an
intra/interspecies interaction constant, $D$ is a diffusion constant, and $%
f_{A,B}(\rho _{A},\rho _{B})$ is a damping function dependent upon the type
of interaction between species $A$, and $B$. It is possible to use these
equations to model either the full counterflowing motion of particles or the
motion of a single species. For instance, if the damping function $f=1$, $%
K_{A,B}<0$, and $D_{A,B}>0$, one obtains a standard diffusion equation for a
single species without interaction:

\begin{equation}
\frac{\partial \rho_{A,B}}{\partial t}=K_{A,B}\frac{\partial \rho_{A,B}}{
\partial x}+D_{A,B}\frac{\partial^2\rho_{A,B}}{\partial x^2}.
\end{equation}
The solution for this equation is given by setting the initial concentration
to a known value $\rho_{A,B}(x,t=0)=g_{A,B}(x)$. On the other hand, the
modeling of two counterflows is given by working with both equations and
setting $K_A\cdot K_B\leq 0$.

As found in some alternatives derived for the interaction between particles%
\cite{rdasilva2015,DZJ2018}, we consider the counterflow of two species in a
directed random walk. In this model, the particles do never return, thus $%
D_A=D_B=0$. Furthermore, the probability of a particle hopping to the next
cell (or remaining in the same cell) is $1/2$ if this cell is empty.
Moreover, this probability decreases in a rate $\alpha$ multiplied by the
ratio between the occupation of one specie and the total occupation of a
cell. Eq. \ref{eq:general_equation} is then rewritten as:

\begin{equation}
\begin{array}{ccc}
\frac{\partial \rho _{A}}{\partial t} & = & -k_{1}\frac{\partial \rho _{A}}{%
\partial x}+k_{2}\frac{\partial }{\partial x}\left( \frac{\rho _{A}\rho _{B}%
}{\rho _{A}+\rho _{B}}\right) \\ 
&  &  \\ 
\frac{\partial \rho _{B}}{\partial t} & = & k_{1}\frac{\partial \rho _{B}}{%
\partial x}-k_{2}\frac{\partial }{\partial x}\left( \frac{\rho _{A}\rho _{B}%
}{\rho _{A}+\rho _{B}}\right)%
\end{array}
\label{eq:coupled}
\end{equation}%
where $k_{1}\geq 0$, $k_{2}\geq 0$. These coupled differential equations
describe interacting counterflowing streams of two oppositely charged
species under a field with magnitude proportional to $\alpha $ in a one
dimensional discretized ring. The damping factor is given in this case by:

\begin{equation}
f_{A,B}(\rho _{A},\rho _{B})=\frac{\rho _{A,B}}{\rho _{A}+\rho _{B}}.
\end{equation}

This problem has been extended for two-dimensional circular crowns\cite%
{Stock2017}. Lanes have been observed in the steady state regime of these
systems exactly as observed in pedestrian dynamics \cite{Pinho2016} or in
the motion of charged colloids\cite{Vissers2011,VissersPRL2011}.
Notwithstanding, such models exclude very important effects such as the self
exclusion of particles. Thus, we have recently proposed a model\cite%
{rdasilva2019} where the occupation of the next cells is given by a modified
Fermi-Dirac distribution. This leads to the following choice for the damping
function:

\begin{equation}
f_{A,B}(\rho_A,\rho_B)=\frac{1}{1+e^{\alpha (\rho_A+\rho_B-\sigma_{\max})}},
\label{Eq:Newf}
\end{equation}%
where $\sigma_{\max}$ works similarly to the Fermi level as the indicator
for the maximum occupation of the cell. This is weighted by the randomness
parameter $\alpha $.

Here we explore the impact of the initial conditions on the TSPDES
considering the choice of $f_{A,B}(\rho _{A},\rho _{B})$ made in Eq. \ref%
{Eq:Newf}. In order to conduct this study we use the Gini coefficient to
investigate the phase transition between a mobile and a clogging phase (or
condensate phase) as partially explored earlier\cite{rdasilva2019}. We show
that the Gini coefficient is capable of indicating important details of this
transition.

We complete this study by performing MC simulations for asynchronous and
synchronous dynamics. We compare both alternatives and also compare them
with numerical solutions of the TSPDES according to the $f_{A,B}(\rho_{A},%
\rho _{B})$ given in Eq. \ref{Eq:Newf}.

In section \ref{Sec:Model} we present the Fermi-Dirac directed random walk
(FDDEW) model as a derivation of the TSPDES to a problem previously defined 
\cite{rdasilva2019}. We obtain recurrence relations for this model and
extend it to the continuous limit.

Section \ref{Sec:Results} covers the solutions of TSPDES where we obtain the
steady state Gini coefficient as function of $\alpha $. We determine a
critical $\alpha _{c}$ coefficient and analyse its dependence upon the
initial distribution of particles $A$ and $B$. We also determine the Gini
coefficient via Monte Carlo (MC) simulations with both synchronous and
asynchronous updating schemes. We finish this work comparing these two
strategies with regards to a mobility parameter previously defined\cite%
{rdasilva2019}.

Section \ref{Sec:Conclusions} concentrates the most important conclusions of
this work.


\section{The model}

\label{Sec:Model}

This work focuses on a two-specie model of particles that drift in
counterflows on an annular system composed of $L$ cells. Considering that
the number of particles (of whichever species) in the following cell affects
the movement of the particles in the present cell, the concentration of the
target objects can be written by the recurrence relation:

\begin{equation}
\rho _{A}(m,n)=p_{m-1,m}^{(n-1)}\rho
_{A}(m-1,n-1)+p_{m,m}^{(n-1)}\rho_{A}(m,n-1)\text{,}
\label{Eq:recurrence_relation_0}
\end{equation}%
where $p_{j,k}$ is the number (or density) of particles of species $A$ in
cell $j$ at time $k$. Moreover, $p_{m,m}^{(n-1)}+p_{m,m+1}^{(n-1)}=1$, since 
$p_{i,j}^{(n)}$ denotes the probability that a particle in cell $i$
(position $x=i\varepsilon $) hops to cell $j$ (position $x=j\varepsilon $)
at $t=$ $n\tau $. In these equations $\tau$ is the time necessary to perform
such transition and $\varepsilon $ is the step length.

Combining these equations, one obtains: 
\begin{equation}
\begin{array}{lll}
\rho _{A}(m,n)-\rho _{A}(m,n-1) & = & p_{m-1,m}^{(n-1)}\rho _{A}(m-1,n-1) \\ 
&  &  \\ 
&  & -\ p_{m,m+1}^{(n-1)}\rho _{A}(m,n-1)%
\end{array}
\label{Eq:recurrence_relation}
\end{equation}

We use a generalized Fermi-Dirac occupation function to model a behavior
similar to that found in solids. The peculiarity of our model is that a
Fermi level-like parameter indicates the maximum occupation of a cell.
Moreover, the inverse temperature relates to a parameter $\alpha$. Thus, the
probability of finding a total of $\sigma_{j,n}=\rho_A(j,n)+\rho_B(j,n)$
particles in cell $j$ at time $n$ is given by:

\begin{equation}
p_{i,j}^{(n)}=\frac{1}{1+e^{\alpha (\sigma _{j,n}-\sigma _{\max })}}.
\label{Eq:FD}
\end{equation}

A process leading to a total number of particles $\sigma_{j,n}>\sigma_{max}$
in a cell should not be as likely as a process that leads to the opposite
result. Also, the likelihood of any of these processes is regulated by $%
\alpha$, which is not necessarily the inverse temperature, but a control
parameter.

Two regimes of this model are of special attention. The first one happens
when $\alpha\rightarrow 0$. In this case the occupation probability $%
p_{i,j}^{(n)}\rightarrow 1/2$, and this is equivalent to a situation where
the particles are small enough so that the interparticle collision is
completely random.

Another limiting situation happens when $\alpha\rightarrow\infty$. This
produces an occupation probability given by:

\begin{equation}
p_{i,j}^{(n)}=\left\{ 
\begin{array}{ll}
1 & \text{if }\sigma _{j,n}=\rho _{A}(j,n)+\rho _{B}(j,n)<\sigma _{\max } \\ 
&  \\ 
1/2 & \text{if }\sigma _{j,n}=\sigma _{\max } \\ 
&  \\ 
0 & \text{otherwise}%
\end{array}%
\right.  \label{Eq:probabilities}
\end{equation}

\begin{figure}[tbh!]
\centering
\includegraphics[width=1.0\columnwidth]{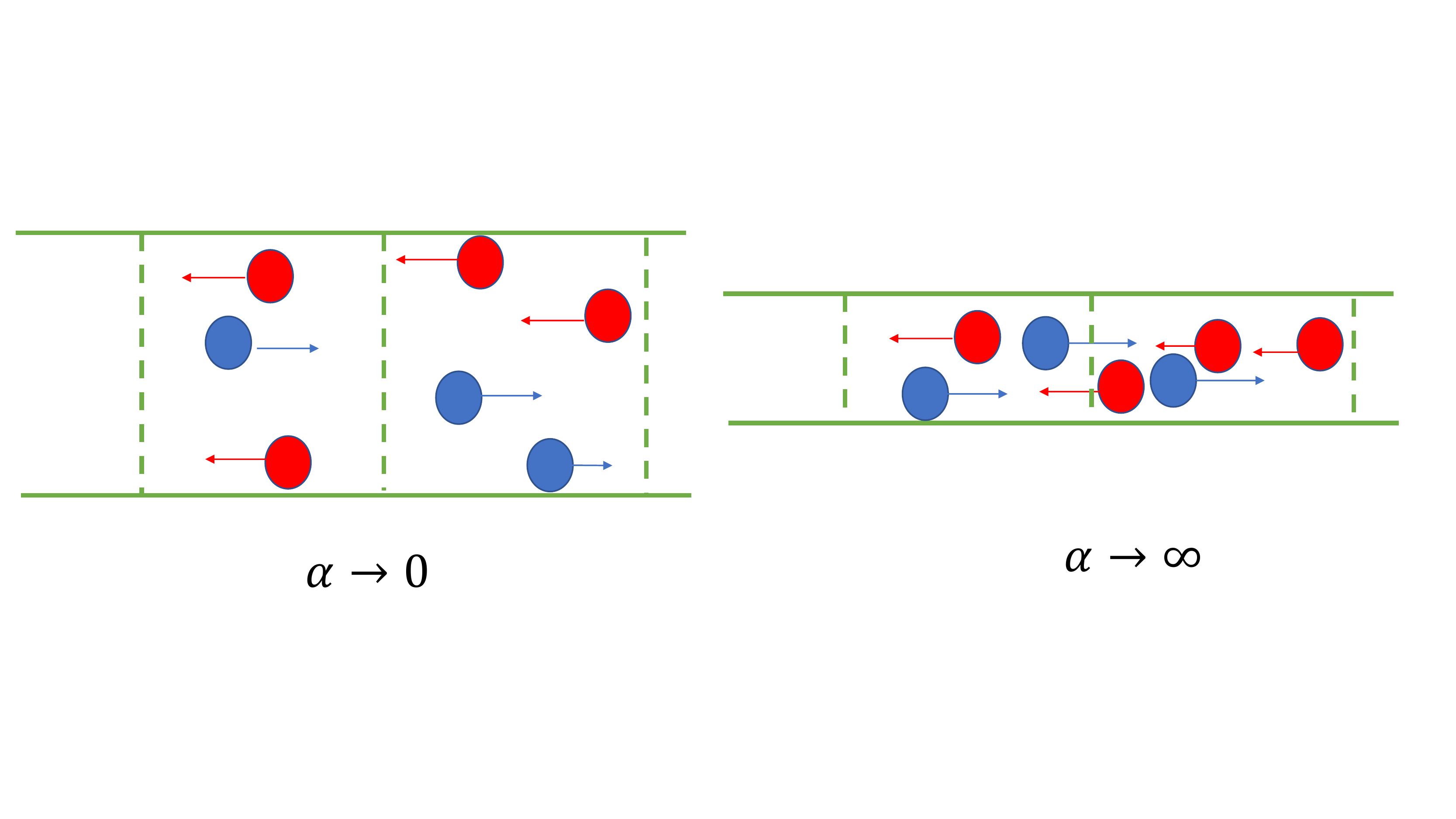}
\caption{Two limiting situations found in our model: a) $\protect\alpha %
\rightarrow 0$ corresponding to small particles with respect to the
dimensions of the cell leading to more random collisions, and b) $\protect%
\alpha \rightarrow \infty $ corresponding to large particles with respect to
the dimensions of the cell leading to more deterministic collisions.}
\label{Fig:scheme}
\end{figure}

This implies that no more than $\sigma _{\max }+1$ objects per cell are
allowed. This corresponds to a situation where particles move in narrow
tube, guided by a field of constant intensity, but due to their large sizes
with respect to the environment, the penetration in the next cells is more
restricted and deterministic. As depicted in Fig. \ref{Fig:scheme} such
dense systems may lead to the formation of condensates.

There is a critical parameter $\alpha_C$ in the intermediate regime $%
\left(\alpha\in(0,\infty)\right)$ where the system loses mobility causing an
abrupt change in the distribution of particles. Setting:

\begin{equation*}
\begin{array}{lll}
a_{m,n} & = & \rho _{A}(m,n)/\left[ 1+e^{\alpha (\rho _{A}(m+1,n)+\rho
_{B}(m+1,n)-\sigma _{\max })}\right] ,\text{and} \\ 
&  &  \\ 
b_{m,n} & = & \rho _{B}(m,n)/[1+e^{\alpha (\rho _{A}(m-1,n)+\rho
_{B}(m-1,n)-\sigma _{\max })}]%
\end{array}%
\end{equation*}%
in Eq. \ref{Eq:recurrence_relation}, one finds:

\begin{equation}
\begin{array}{lll}
\rho _{A}(m,n) & = & \rho _{A}(m,n-1)+a_{m-1,n-1}-a_{m,n-1} \\ 
&  &  \\ 
\rho _{B}(m,n) & = & \rho _{B}(m,n-1)+b_{m+1,n-1}-b_{m,n-1}%
\end{array}%
\end{equation}

Combining the terms, and taking the continuous limit one then obtains: 
\begin{equation}
\frac{\partial \rho _{A(B)}(x,t)}{\partial t}=-(+)C\frac{\partial }{\partial
x}\left[ \frac{\rho _{A(B)}(x,t)}{1+e^{\alpha (\rho
_{A}(x,t)+\rho_{B}(x,t)-\sigma_{\max})}}\right],  \label{Eq:EDP}
\end{equation}
where $C=\lim_{\tau ,\varepsilon \rightarrow 0}\frac{\varepsilon }{\tau }$
corresponds to a particular case of Eq. \ref{eq:general_equation} for the
damping factor shown in Eq. \ref{Eq:Newf}.


\subsection{Monte Carlo simulations}

It is possible to either solve Eq. \ref{Eq:EDP} directly or via Monte Carlo
simulations. For the latter, we consider that a particle of species $A$ at
instant $i$ in cell $j$ will occupy the cell $j+1$ with probability $%
p_{i,j+1}^{(n)}=\frac{1}{1+e^{\alpha(\sigma _{j,n}-\sigma _{\max })}}$.

This simulation can be performed with two different updating schemes:

\begin{enumerate}
\item \textbf{Synchronous update}: All particles are simultaneously updated.
This choice was performed in \cite{rdasilva2019};

\item \textbf{Asynchronous update}: We select $n_{part}$ particles (the
total number of the particles) distributed in $L$ cells. The update is
performed for each particle leaving the remaining ones unchanged.
\end{enumerate}

In order to quantify the fluctuations and find the transition between a
mobile and a clogged phase we propose the use of the Gini coefficient $G(t)$%
\cite{Gini1921}. Since the number $\rho _{A,B}(j,n)$ of particles from
species $A$ and $B$ at cell $j=1,...,L$ at an instant $n$ is known, the Gini
coefficient for the concentration of particles is given by:

\begin{equation}
G(n)=\frac{1}{(L-1)}\left[ L+1-2\left( \frac{\sum_{j=1}^{L}(L+1-j)%
\rho_{A,B}(j,n)}{\sum_{j=1}^{N}\rho_{A,B}(j,n)}\right) \right] .
\label{Eq:Gini}
\end{equation}

Unlike other order parameters such as the mobility\cite{rdasilva2019}, the
Gini coefficient can be calculated during both Monte Carlo simulations and
the direct solution of the TSPDES. Nonetheless, the mobility is given by:

\begin{equation}
M(n) = \frac{1}{N_{part}}\sum_{i=1}^{N_{part}}\xi _{i}(n),
\label{Eq:mobility}
\end{equation}%
where $\xi _{i}(n)$ is a binary variable associated to particle $i$ that
assumes 0 if the particle stays still at time $n$ and 1 if this same
particle hops to the next cell at that period. This quantity cannot be
calculated from the solution of the recurrent relations. Nonetheless, it is
easily obtained from the MC simulations since they are performed directly on
the particles.

It is important to notice that unlike the direct solutions of TSPDES, the
steady state values of $G_{\infty }$ and $M_{\infty }$ for MC simulations
depend only slightly on the run showing little ensemble variability.
Furthermore, the time $t_{steady}$ necessary to reach the steady state
regime depends on the value of $\alpha $.

This stochastic behavior was not observed when we directly solved the
TSPDES. In this case $10^{5}$ iterations are enough to obtain $G_{\infty}$
and $M_{\infty}$. The slopes of linear fittings for the last $\Delta =10^3$
values of $G$ and $M$ were calculated for MC simulations until both of them
were smaller than $\eta =10^{-7}$.

In the next section we present the main results obtained from both the
direct solution of the TSPDES and MC simulations.


\section{Results}

\label{Sec:Results}

We studied the time evolution of particles of species $A$ counterflowing
particles of species $B$ under three distinct initial conditions:

\begin{enumerate}
\item \textbf{Dirac Delta Pulses (DDP)}: All particles of species $A$ are
placed in the cell $i$ and all particles of species $B$ are placed in cell $%
i+L-1$, where $L$ is the number of cells in the simulation;

\item \textbf{Uniform Distribution (UD)}: The same number of particles of
species $A$ and $B$ are uniformly distributed over the $L$ cells;

\item \textbf{Constant Occupation (CO)}: Each cell has two particles: one of
species $A$ and another of species $B$.
\end{enumerate}


\subsection{DDP Initial Conditions}

We simulated a ring with $L=128$ cells by integrating the TSPDES. The DDP
initial conditions in this case are explicitly given by $\rho
_{A}(x,t=0)=L\delta _{x,0}$ and $\rho _{B}(x,t=0)=L\delta _{x,L}$, with:%
\begin{equation}
\delta _{x,y}=\left\{ 
\begin{array}{ccc}
1 & \text{se} & x=y \\ 
&  &  \\ 
0 & \text{se} & x\neq y%
\end{array}
\right.
\end{equation}

Figures \ref{Fig:delta_alfa=0,4_time_evolution}, \ref%
{Fig:delta_alfa=3,0_time_evolution}, and \ref%
{Fig:delta_alfa=20,0_time_evolution} show the distributions of concentration
for $\alpha =0.4$, $3.0$, and $20.0$ respectively for different periods ($%
t=0,10^3,10^4$ and $10^5$).

\begin{figure}[tbh!]
\centering
\includegraphics[width=0.7%
\columnwidth]{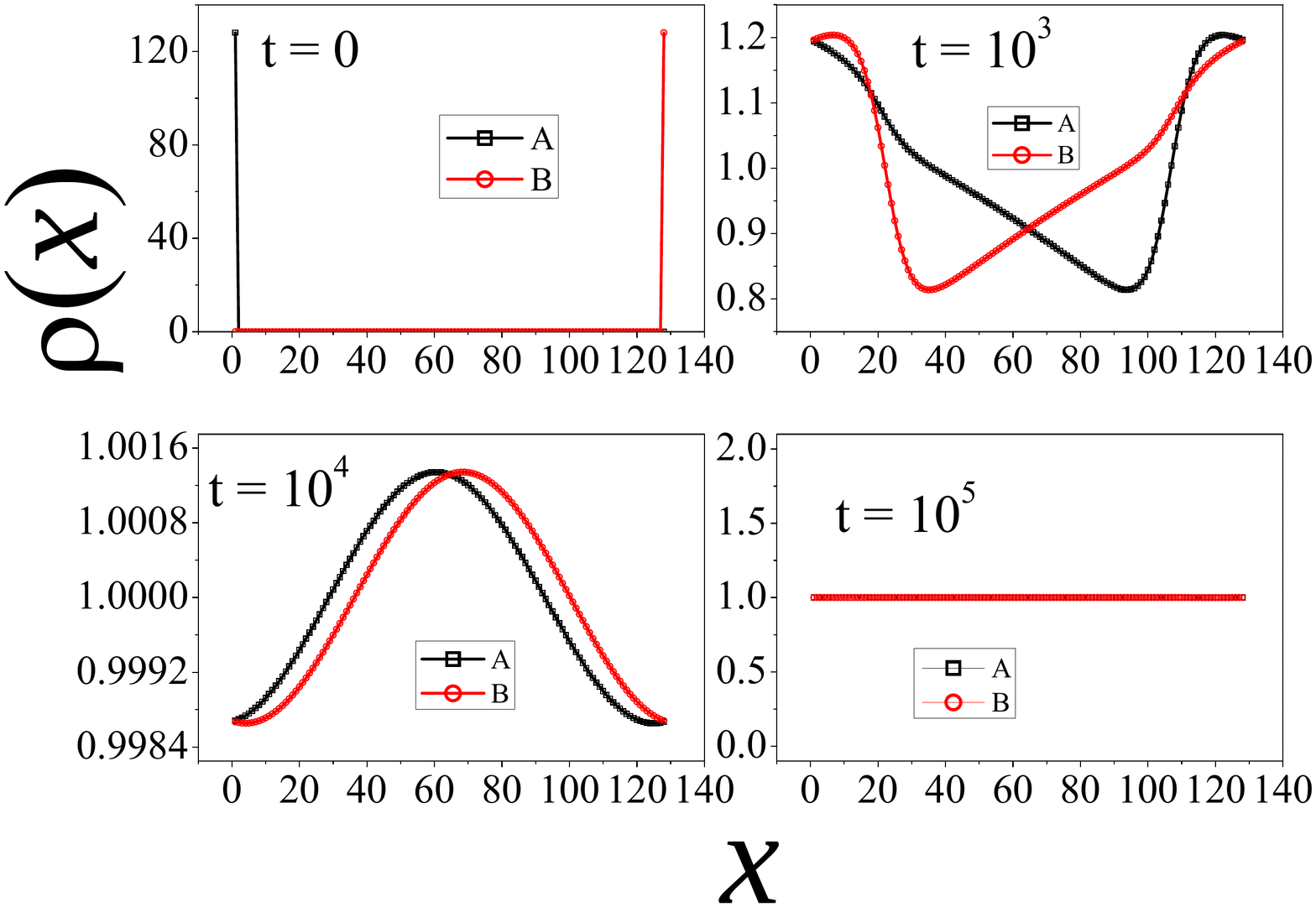}
\caption{Temporal evolution of the densities of particles under DDP initial
conditions and $\protect\alpha=0.4$. Steady state is reached for $t>10^{4}$
when particles flow without clogging effects, i.e., $\protect\rho _{A}=%
\protect\rho _{B}=1$ for all cells. }
\label{Fig:delta_alfa=0,4_time_evolution}
\end{figure}

For $\alpha=0.4$, the steady state regime is reached when $t>10^4$. In this
situation the particles flow without any clogging effects and are uniformly
distributed along the ring, i.e., $\rho_A=\rho_B=1$ for all cells. This
behavior is shown in Fig. \ref{Fig:delta_alfa=0,4_time_evolution}.

\begin{figure}[tbh]
\centering
\includegraphics[width=0.7%
\columnwidth]{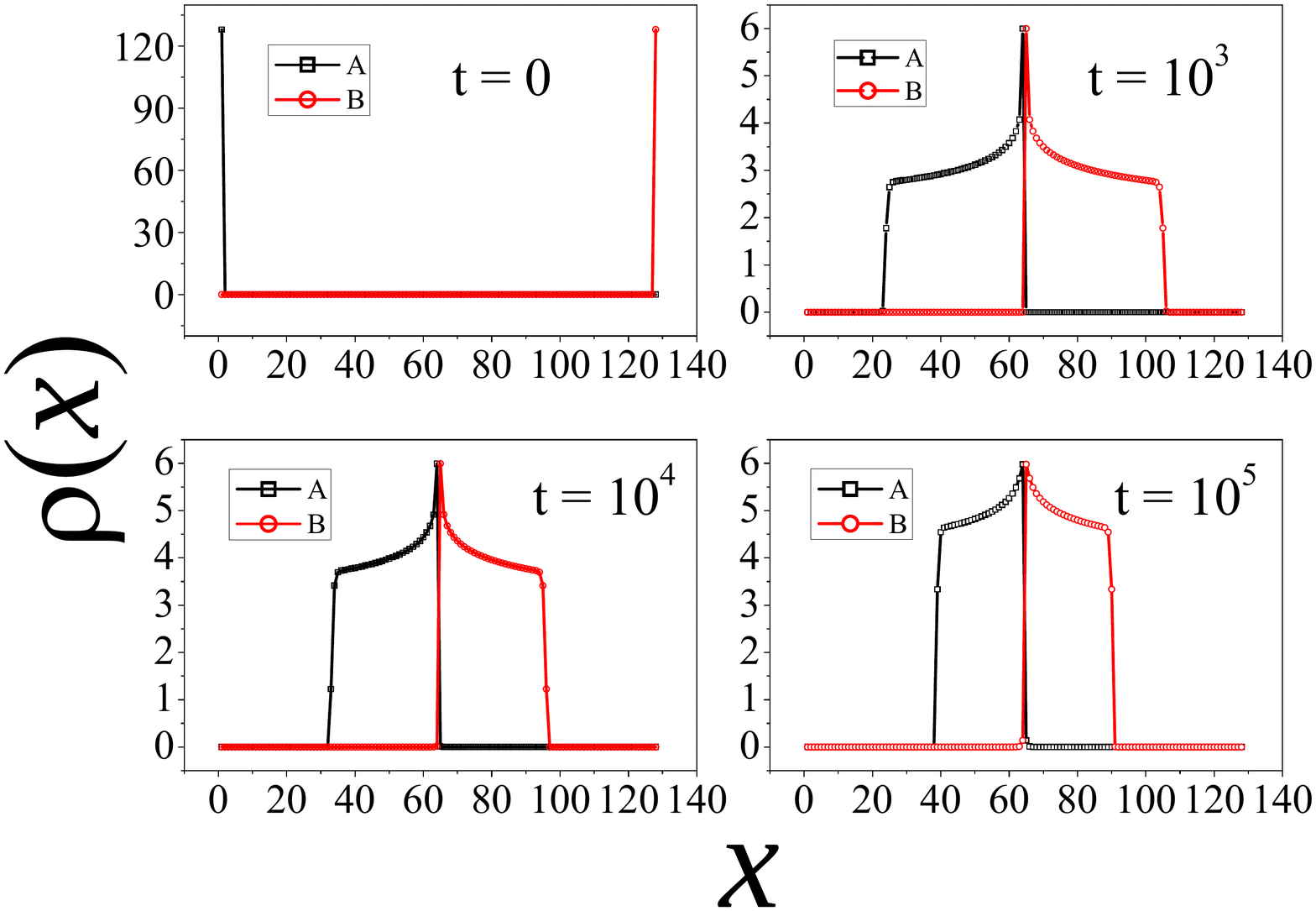}
\caption{Temporal evolution of the densities of particles under DDP initial
conditions and $\protect\alpha=3$, representing a higher level of disorder.
All other parameters are the same as used in Fig. \protect\ref%
{Fig:delta_alfa=0,4_time_evolution}. Here it is possible to observe the
formation of condensates (peaks in density) along the ring.}
\label{Fig:delta_alfa=3,0_time_evolution}
\end{figure}

The onset of condensation happens at $\alpha=3$ as depicted in Fig. \ref%
{Fig:delta_alfa=3,0_time_evolution}. At this level of disorder it is
possible to observe peaks of approximately 6 particles in a cell. Complete
clogging is obtained at $\alpha=20$ as shown in Fig. \ref%
{Fig:delta_alfa=3,0_time_evolution}. In this case, clogging happens when the
two species meet for the first time.

\begin{figure}[tbh!]
\centering
\includegraphics[width=0.7%
\columnwidth]{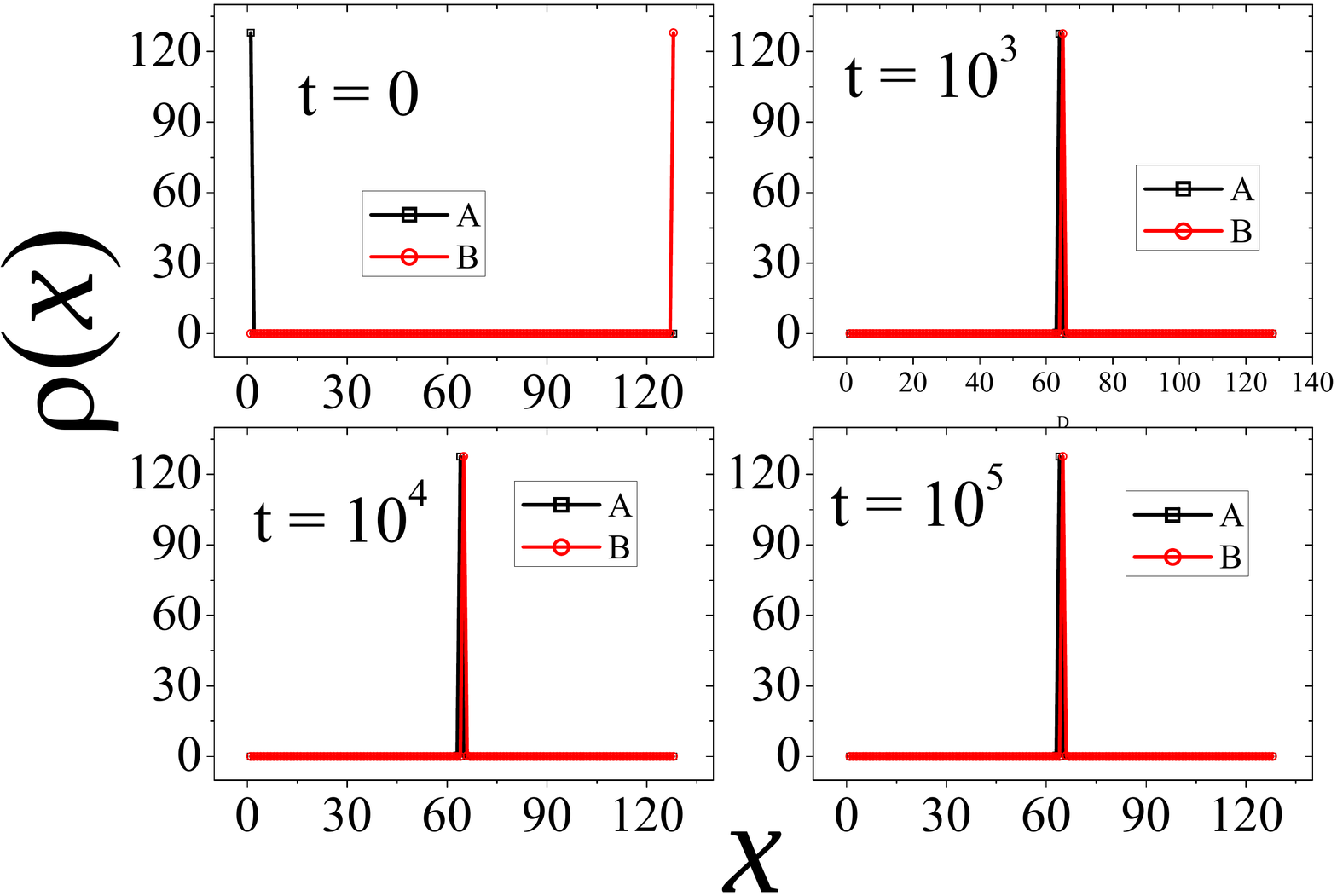}
\caption{Density of particles for $\protect\alpha =20$. The particles enter
a complete clogging state after meeting for the first time. Both
concentrations display a central peak at around 128 particles in a single
cell.}
\label{Fig:delta_alfa=20,0_time_evolution}
\end{figure}


\subsection{UD Initial Conditions}

The steady state regime shows a peculiar behavior for UD initial conditions.
Here we analyzed the distributions for three distinct values of $\alpha$ ($%
0.45, 0.55,$ and $10$). These results are shown in Fig. \ref%
{Fig:time_evolutions_from_uniform} a, b, and c.

\begin{figure}[tbh!]
\centering
\includegraphics[width=0.52%
\columnwidth]{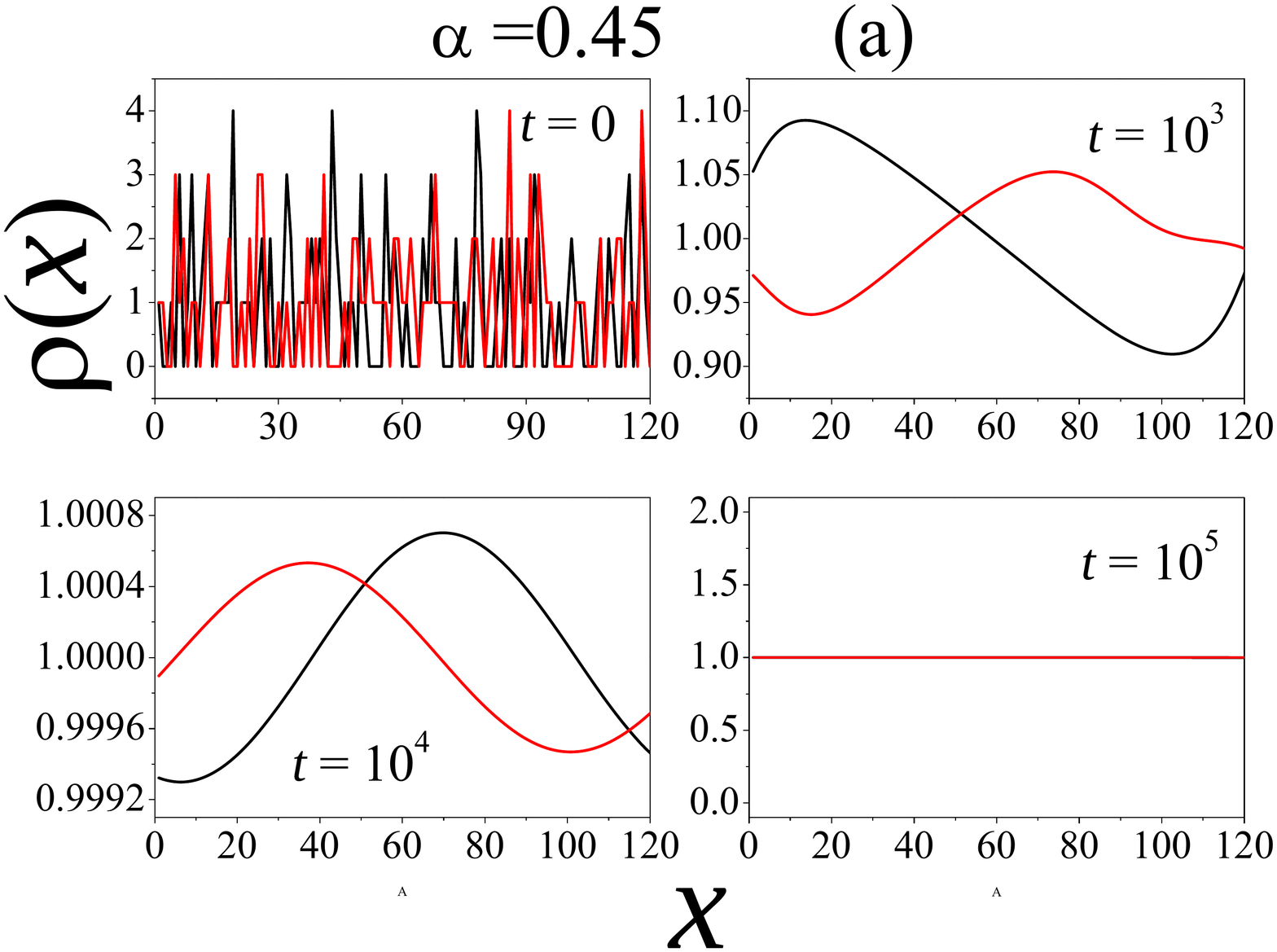} %
\includegraphics[width=0.52%
\columnwidth]{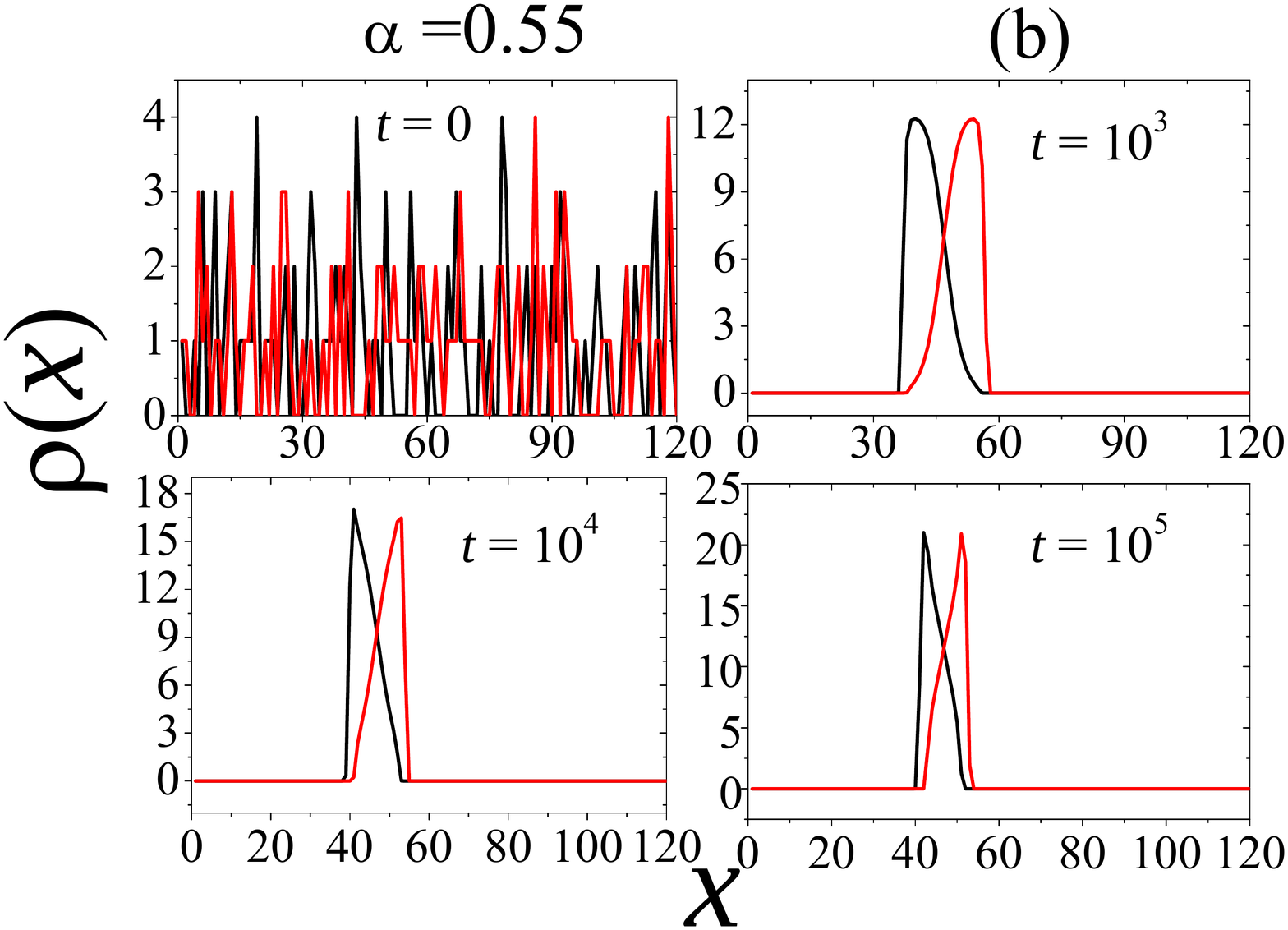} %
\includegraphics[width=0.52%
\columnwidth]{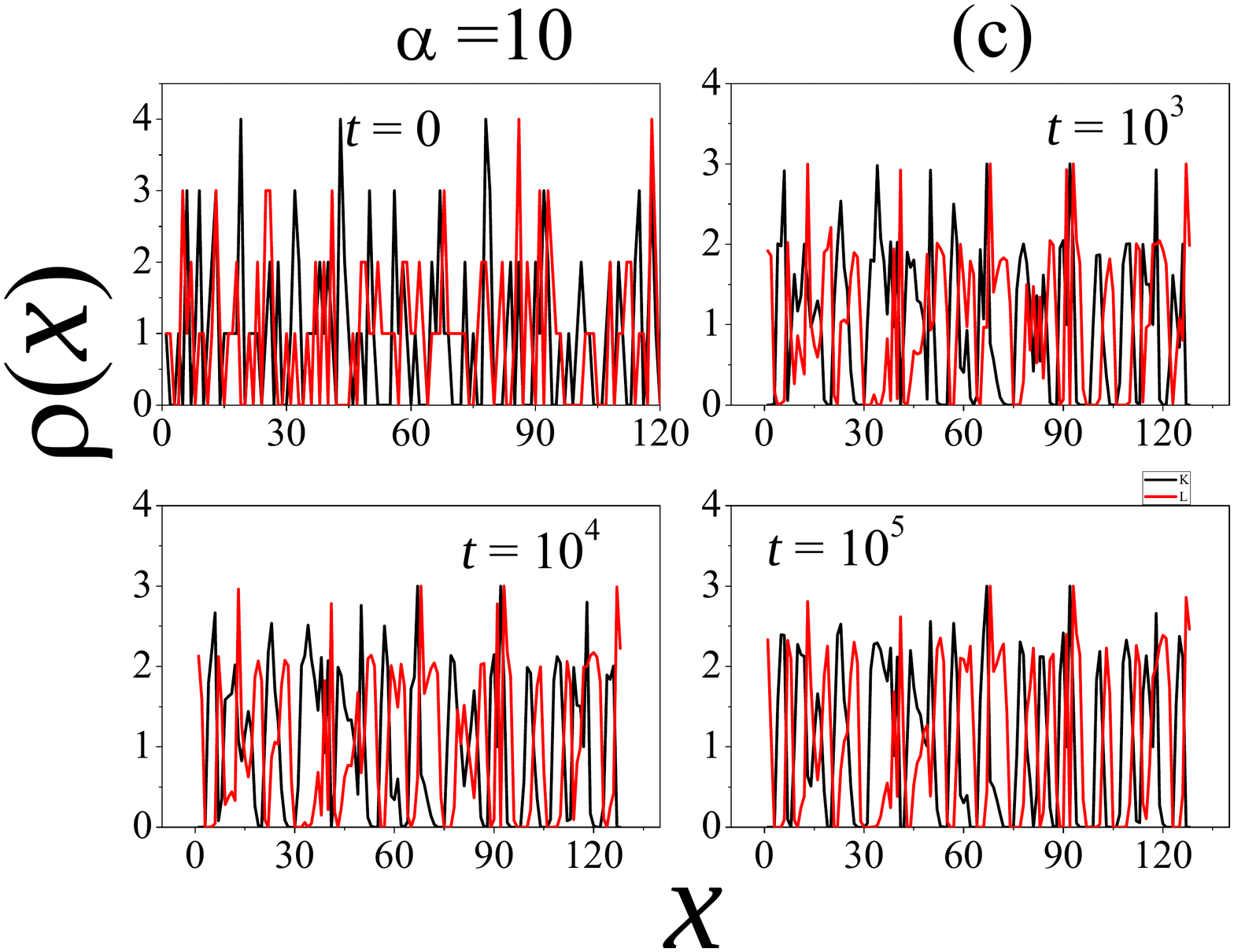}
\caption{Density of particles for a) $\protect\alpha=0.45$, b) $0.55$, and
c) $10$. The initial condition corresponding to uniformly distributed
particles is the same for all cases.}
\label{Fig:time_evolutions_from_uniform}
\end{figure}

It is possible to observe in Fig. \ref{Fig:time_evolutions_from_uniform}a
that the system evolves to a completely mobile phase without condensation.
As depicted by a pronounced peak in Fig. \ref%
{Fig:time_evolutions_from_uniform}b, condensation appears when $\alpha=0.55$%
. This characterizes a clogging phase.

The system then moves to a state composed of small condensates for values of 
$\alpha$ as high as 10. This can be observed in Fig. \ref%
{Fig:time_evolutions_from_uniform}c. Unlike to pronounced peak found for $%
\alpha=0.55$, here the particles are uniformly distributed.


\subsection{Gini Coefficient}

The collective phenomena can be further accessed by the Gini coefficient.
Since it measures the heterogeneity of the distribution, we use it to
quantify the condensation of the system. For instance, a single peak
corresponds to a maximum Gini coefficient ($G\rightarrow 1$). A distribution
of condensates yields intermediate values ($G\sim 1/2$). Finally, a
homogeneous distribution corresponding to the absence of condensation
produces small Gini values ($G\rightarrow 0$).

Simulations based on the integration of the TSPDES were performed for $%
t_{\infty }=10^{4}$ iterations. This is sufficient to reach a steady state
value for $G_\infty$. On the other hand, it was not possible to achieve
convergence for a single $G_\infty$ value in MC simulations. We used $%
G_\infty\ \text{or}\ M_{\infty}<10^{-7}$ as a stop condition.

In order to verify whether $G_\infty$ is a good estimator for the
clogging-mobile transitions, it was measured as a function of $\alpha$ for 
\emph{i}) asynchronous MC simulations, \emph{ii}) synchronous MC
simulations, and \emph{iii}) numerical TSPDES integration.

\begin{figure}[tbh!]
\centering
\includegraphics[width=1.0\columnwidth]{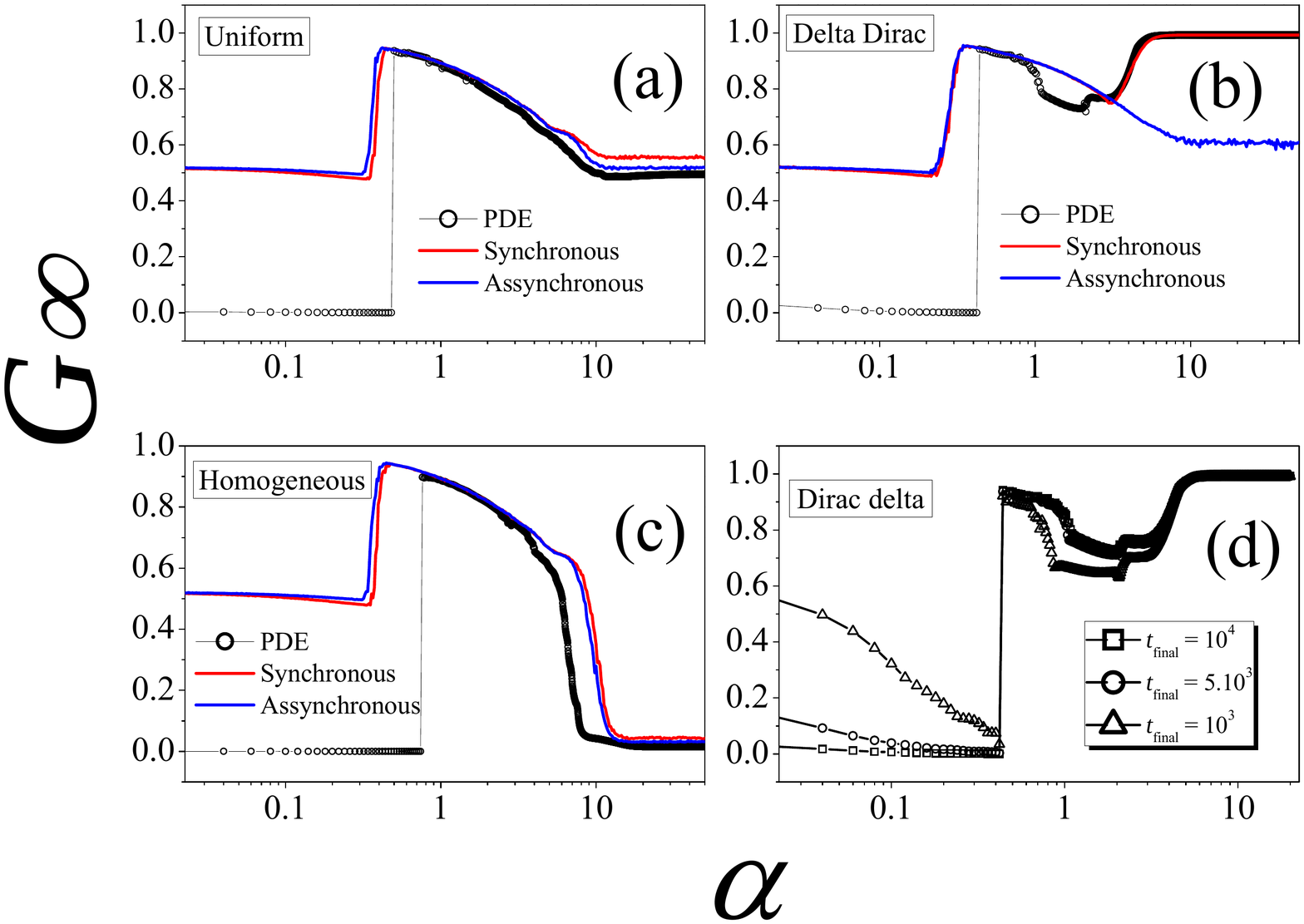}
\caption{Steady state Gini coeficient as function of $\protect\alpha$ for: 
\emph{i}) the numerical integration of the TSPDES, \emph{ii}) asynchronous
MC simulations, and \emph{iii}) synchronous MC simulations. Results for UD,
DDP, and CO initial conditions are shown in (a), (b), and (c), respectively.
The effects of the integration period is shown in (d).}
\label{Fig:stationary_gini}
\end{figure}

The Gini coefficient changes abruptly as a function of $\alpha$ as shown in
Figs. \ref{Fig:stationary_gini} (a), (b), and (c). Although this transition
does not occurr exactly at the same value of $\alpha$, there is a
qualitative agreement between the MC simulations and the numerical
integration of the TSPDES. Moreover, the transition always occurr for $%
\alpha<1$.

The Gini coefficient tends to 0 for the numerical integration of the TSPDES
and $1/2$ for MC simulations in the mobile phase (low $\alpha$). The latter
case corresponds to a situation where there are two types of occupation of
the cells, whereas the former corresponds to a homogeneous occupation. Thus,
it is possible to state that MC simulations produce \emph{statistically
homogeneous steady state regimes}, whereas the numerical integration of the
TSPDES produces \emph{completely homogeneous steady state regimes}.

\begin{figure}[tbh!]
\centering
\includegraphics[width=0.6\columnwidth]{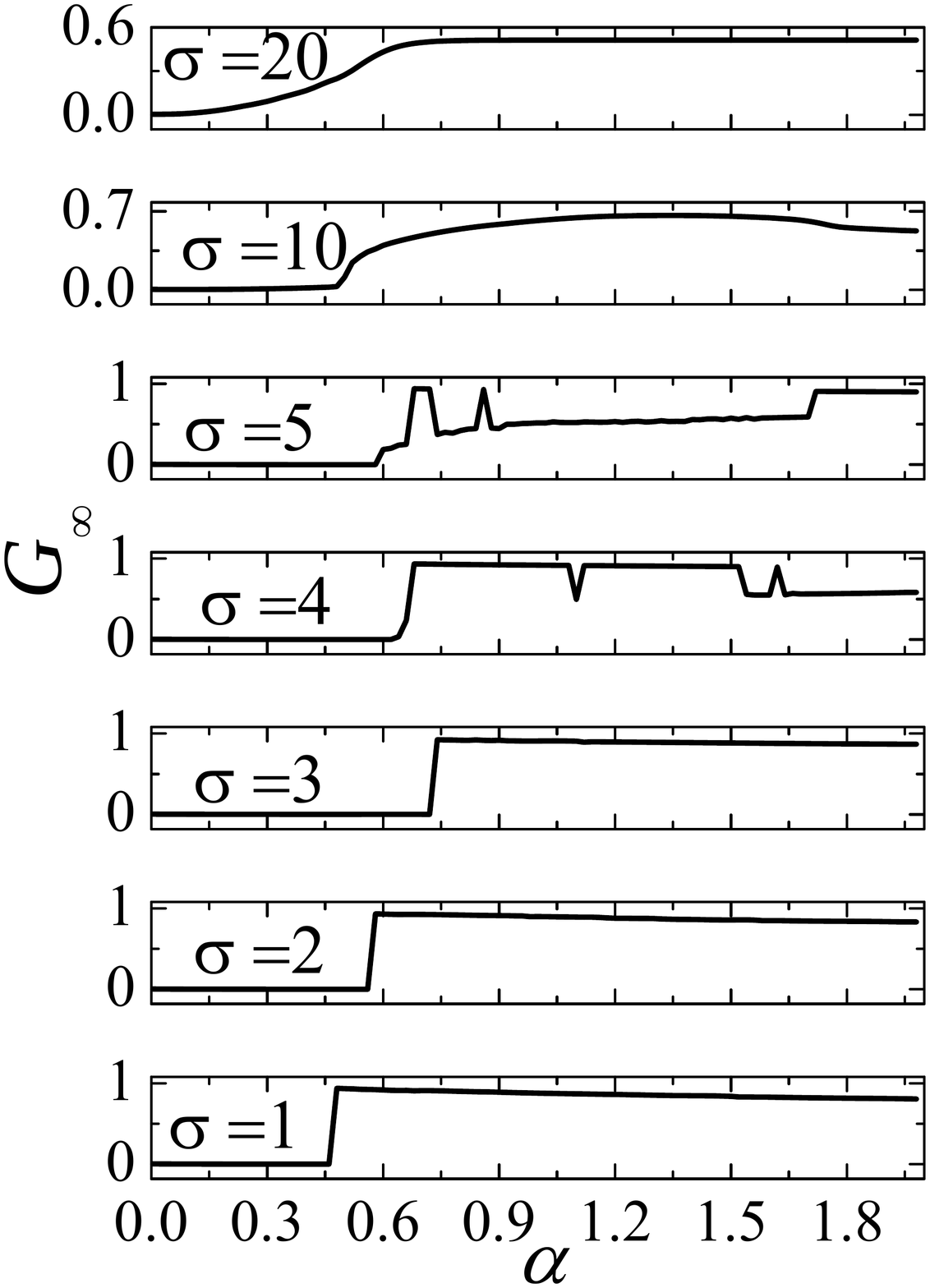}
\caption{$G_\infty$ as a function of $\protect\alpha$ obtained from the
numerical integration of the TSPDES for different values of $\protect\sigma%
_{max}$. In all cases $L=128$ and $N_{part}=128$ for each species. The
initial condition was a uniform distribution of particles (UD).}
\label{Fig:stationary_gini_density_constant}
\end{figure}

Initial uniform distributions do not change in time if $\alpha$ is
sufficiently high. Thus, $G_{\infty}\approx 1/2$ for both MC simulations and
the numerical integration as depicted in Fig. \ref{Fig:stationary_gini}a.
Fig. \ref{Fig:stationary_gini}b makes explicit the wave behavior of the model.
The system departs from two concentrated groups and produces a transition
simultaneously for both synchronous MC simulation and TSPDES integration.
This happens because of the synchronicity of both schemes. After the two
groups collide they produce a permanent condensation that does not
dissipate. On the other hand, asynchronous MC simulations have a diffusive
character. Hence, particles are dispersed and the condensation is less
strong than found in the other cases. Therefore, the Gini coefficient is
smaller. Furthermore, both the synchronous and asynchronous MC simulations
show a perfect agreement in the limit of low interaction for $\alpha < 3$.

The Gini coefficient for a homogeneous initial distribution is shown in Fig. %
\ref{Fig:stationary_gini}. Unlike the behavior described for uniform and
concentrated initial conditions, $G_\infty\rightarrow 0$ for high values of $%
\alpha$ when the system departs from a homogeneous occupation. This happens
because the system does not change its state when $\alpha$ is too high. A
numerical strategy was used to perform the numerical integration in this
situation. The center site was initially set to zero particles, whereas the
remaining ones were set to one particle: $\rho_{A,B}(x,t=0)=1-%
\delta_{x,L/2} $.

The integration period $t_{\infty }$ is particularly important for the
numerical integration of the TSPDES. Fig. \ref{Fig:stationary_gini}d shows
the Gini coefficient calculated for $t_{\infty }=1.0\times 10^{3},5.0\times
10^{3},\ \text{and }1.0\times 10^{4}$. Since there is no considerable
differences between the Gini coefficients calculated for $t_{\infty
}=5.0\times 10^{3},\ \text{and }1.0\times 10^{4}$, one can assume that the
latter is a suitable value for the simulations.


\subsection{Occupation Effects and Dilution}

In order to study dilution and occupation effects, we integrated the TSPDES
for a set of $N_{part}=128$ particles of each species uniformly distributed
over $L=128$ cells. $G_\infty$ as a function of $\alpha$ was monitored while 
$\sigma_{\max}$ was varied from 1 to 20.

As Fig. \ref{Fig:stationary_gini_density_constant} shows, the critical
transition initially moves to higher values of $\alpha$ as $\sigma$
increases. When $\sigma=4$ this trend ends and the transitions becomes
smoother, often showing irregularities with an onset that reduces with an
increasing $\sigma$. This is a consequence of the system having more degrees
of freedom because of larger capacities of the cells. Thus, the system
becomes more mobile. This new trend continues until $\sigma_{\max}=20$. At
this stage, the transition becomes continuous, its order changes, and $%
G_\infty$ reduces.

Here we used a constant density of particles of each species: $\overline{\rho%
}=N_{part}/L=N_0$. Another way of carrying this simulation is by keeping the 
\emph{occupation} constant. This is defined by $C=N_{part}/(\sigma_{%
\max}L)=C_0$. Thus, $L$ remains constant, but $\sigma_{\max}$ and $L_{part}$
can be varied during the simulation.

\begin{figure}[tbh!]
\centering
\includegraphics[width=0.7%
\columnwidth]{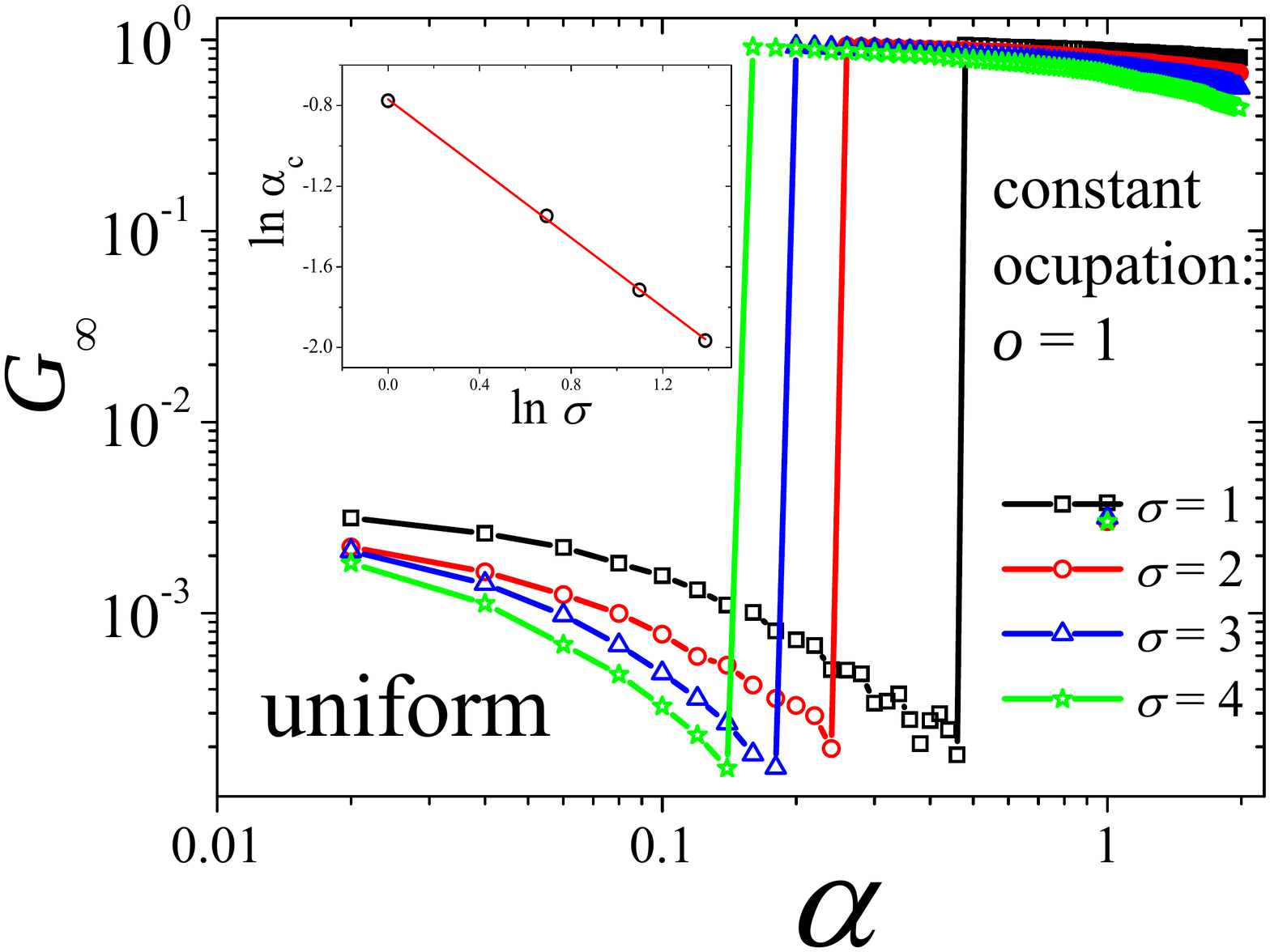}
\caption{$G_{\infty}$ as function of $\protect\alpha $ obtained from the
numerical integration of TSPDES for different values of $\protect\sigma%
_{\max}$. In all cases, $L=128$ and $N_{part}$ is varied to keep $%
C=N_{part}/(L\protect\sigma_{\max})$ constant. The particles of each species
were initially uniformly distributed over the cells (UD).}
\label{Fig:stationary_gini_ocupation_constant}
\end{figure}

Fig. \ref{Fig:stationary_gini_ocupation_constant} shows the behavior of $%
G_\infty$ as a function of $\alpha$. As the insect indicates, the critical
value of $\alpha$ has a power law dependence with $\sigma_{\max}$:

\begin{equation}
\alpha_{c}\sim \sigma_{\max}^{-\Delta},
\end{equation}
where $\Delta \approx 0.86$.


\subsection{Finite Size Scaling and Effects of the Updating Schemes on the
Mobility}

\begin{figure}[tbh!]
\centering
\includegraphics[width=0.7%
\columnwidth]{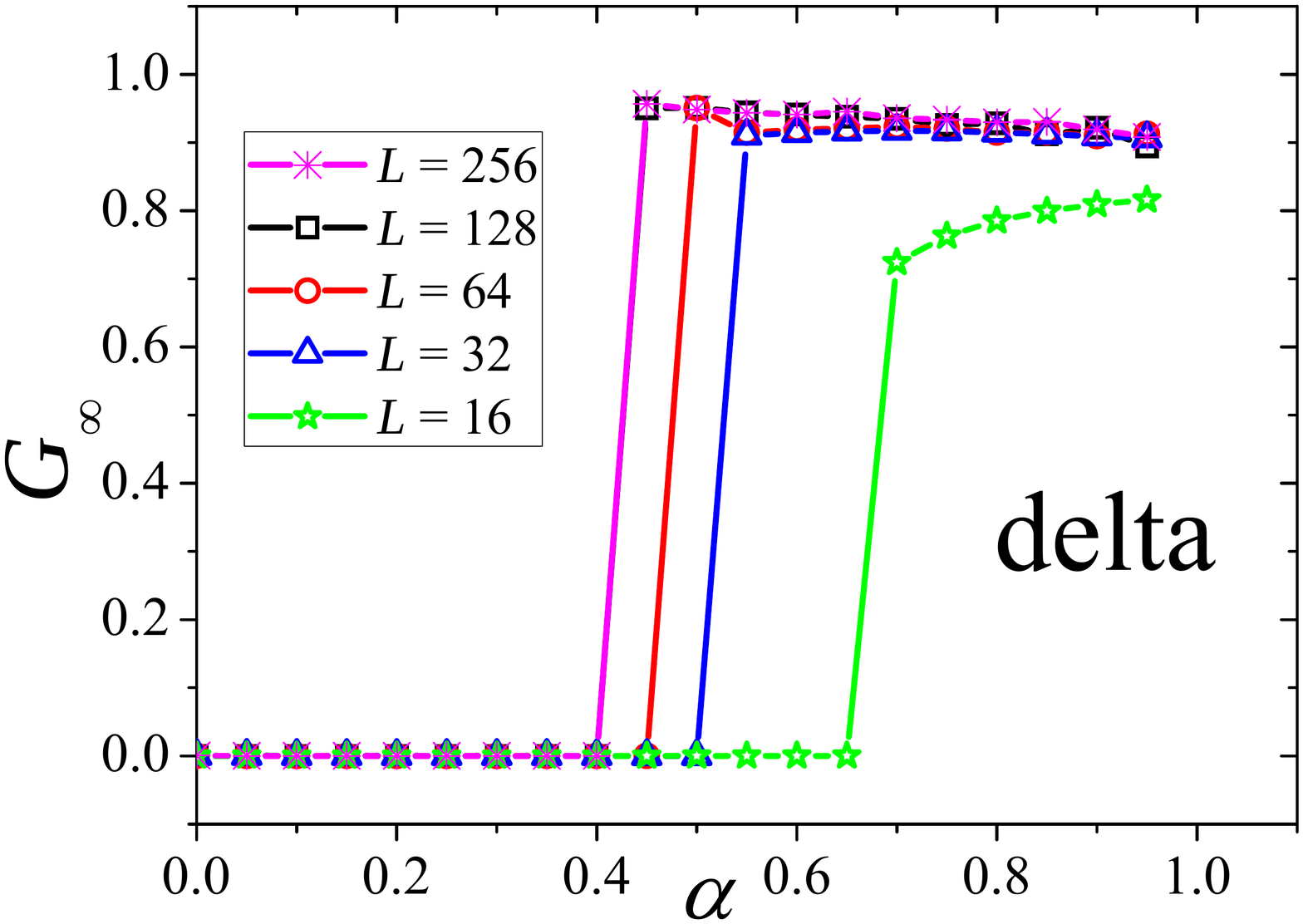}
\caption{Finite size scaling considering DDP initial conditions. No
significant variations can be observed for $L>128$.}
\label{Fig:finite}
\end{figure}

$G_\infty$ was measured as a function of $\alpha$ for different values of $L$
using DDP initial conditions. The results shown in Fig. \ref{Fig:finite}
indicate that no significant variations are observed for $L>128$. Therefore, 
$L=128$ is an appropriate value for our simulations.

Finally, the steady state mobility $M_{\infty }$ for synchronous \cite%
{rdasilva2019} and asynchronous updating schemes were compared.

\begin{figure}[tbh!]
\centering
\includegraphics[width=0.6\columnwidth]{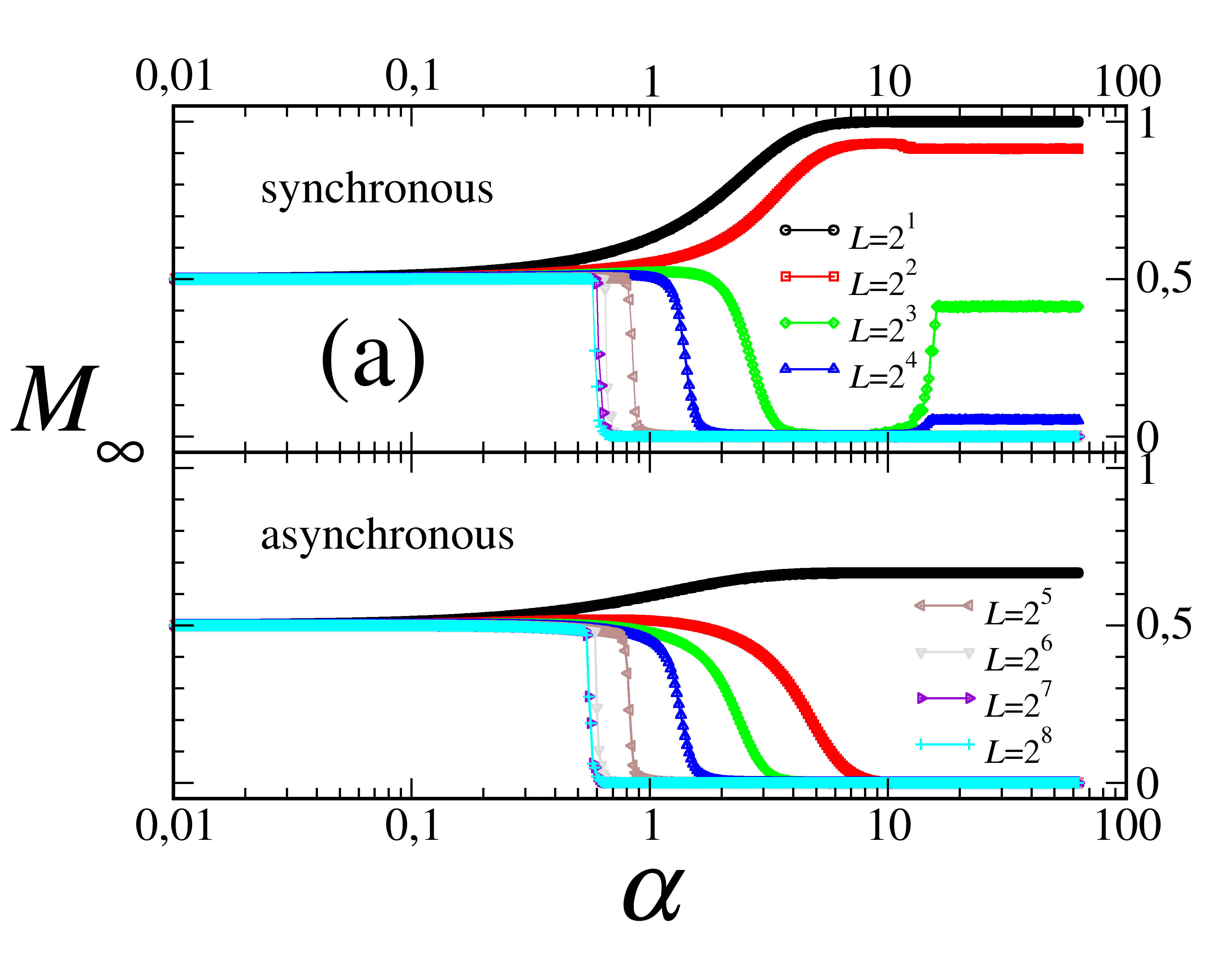} %
\includegraphics[width=0.6%
\columnwidth]{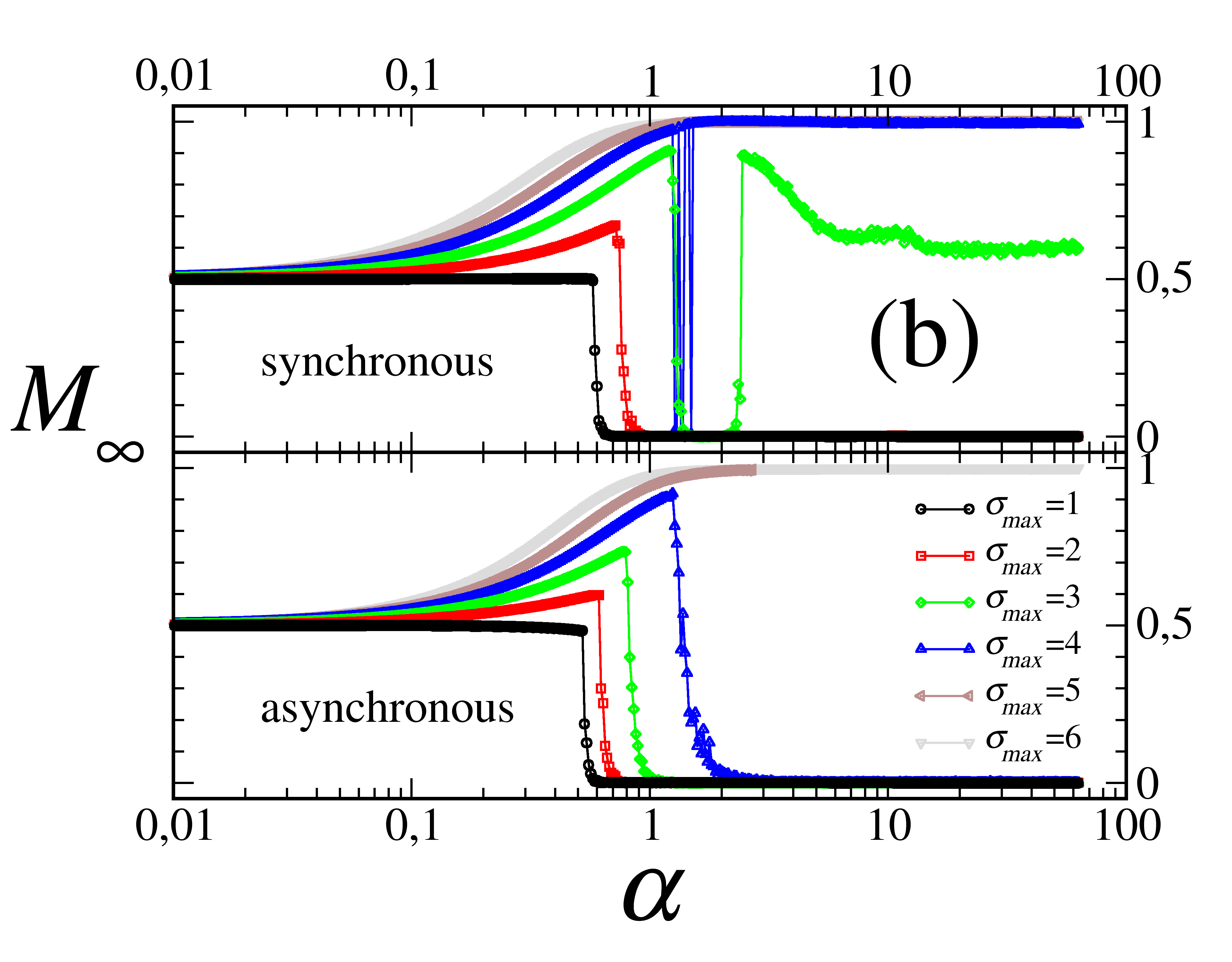} %
\includegraphics[width=0.6%
\columnwidth]{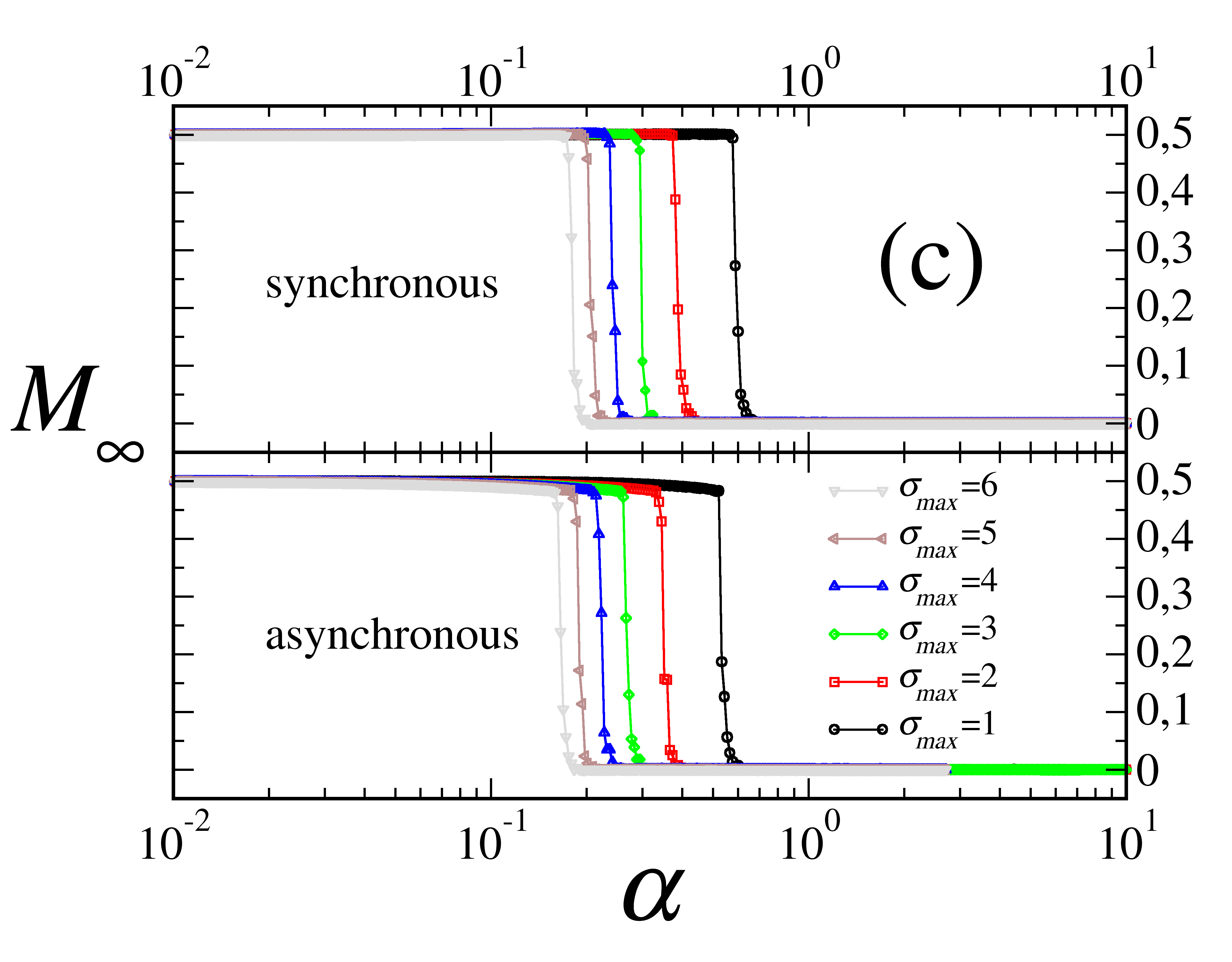}
\caption{The effects of synchrounous and asynchronous updating schemes on
the steady state mobility varying: a) the system size, b) $\protect\sigma%
_{\max}$ while keeping the density $\protect\rho=N_{part}/L=1$ constant, and
c) $\protect\sigma_{\max}$ while keeping the occupation $C=N_{part}/(\protect%
\sigma_{\max}L)=1$ constant. $L=128$ for all cases.}
\label{Fig:mobility}
\end{figure}

Fig. \ref{Fig:mobility}a shows that a transition between a clogged
(condensation) and a mobile phase can be observed for large systems $%
L\geq2^4 $. As the system is made bigger, the transition becomes more
abrupt. Small systems, though, show false mobile phases even for large
values of $\alpha$.

This anomaly is more pronounced in the synchronous dynamics, where the
system can recover its mobility for very large values of $\alpha$. This is
explained by the symmetry of the Fermi-Dirac distribution. Nonetheless, this
effect is severely reduced for $L>2^4$, which is still a small system. This
is further explained elsewhere \cite{rdasilva2019}. On the other hand, this
is not observed for the asynchronous dynamics even for extremely small
systems.


\section{Summary and conclusions}

\label{Sec:Conclusions}

This works extends a previous study on the synchronous dynamics of
counterflowing particles.\cite{rdasilva2019} Here we studied its
asynchronous Monte Carlo dynamics and expanded the analysis for different
initial conditions. Our results indicate that there is a transition from a
mobile to a condensation phase. Furthermore we used the Gini coefficient as
a non-conventional order parameter.

We began our discussion showing that the problem of counterflowing streams
of particles is more general than previously stated. Moreover, we show that
the Fermi-Dirac directed random walk used in this study is appropriate to
model the clogging-mobile transition.

Furthermore, the level of randomness (or determinism) of the system is
determined by the ratio between the size of the particles and the thickness
of the tube wherein they flow. Thus, our study suggests that the steady
state properties of the system are strongly dependent upon the initial
conditions at which the system is prepared.


\section*{Acknowledgements}

R. da Silva and E. V. Stock were financially supported by CNPq under grant
numbers: 311236/2018-9, 424052/2018-0, and 154822/2016-7. This work was
partly developed using the resources of Cluster Ada, IF-UFRGS.



\section*{References}

\begin{thebibliography}{99}
\bibitem{Schmittmann1992} B. Schmittmann, N. Hwang, R. K. P. Zia, \emph{%
Europhys. Lett.} \textbf{19}, 19-25 (1992).

\bibitem{Helbing2000} D. Helbing, I. J. Farkas, T. Vicsek, \emph{Phys. Rev.
Lett.} \textbf{84}, 1240 (2000).

\bibitem{Marroquin2014} F. Alonso-Marroqu\i n, J. Busch, C. Chiew, C.
Lozano, A. Ram\i rez-Gomez, \emph{Phys. Rev. E} \textbf{90}, 063305 (2014).

\bibitem{Vissers2011} T. Vissers, A. Wysocki, M. Rex, H. Lowen, C. P.
Royall, A. Imhof, A. van Blaaderen, \emph{Soft Matter}, \textbf{7}, 2352
(2011).

\bibitem{VissersPRL2011} T. Vissers, A. van Blaaderen, A. Imhof, \emph{Phys.
Rev. Lett.} \textbf{106}, 228303 (2011).

\bibitem{Pinho2016} C. L. N. Oliveira, A. P. Vieira, D. Helbing, J. S.
Andrade Jr., H. J. Herrmann, \emph{Phys. Rev. X}, \textbf{6} 011003 (2016).

\bibitem{Stock2017} E. V. Stock, R. da Silva, and H. A. Fernandes, \emph{%
Phys. Rev. E} \textbf{96}, 012155 (2017).

\bibitem{Wei2015} J. Wei, H. Zhang, Y. Guo, M. Gu, \emph{Phys.} \emph{Lett. A%
} \textbf{379}, 1081--1086 (2015).

\bibitem{Majundar2005} S. N. Majundar, M. R. Evans, R. L. P. Zia, \emph{%
Phys. Rev. Lett.} \textbf{94}, 180601 (2005).

\bibitem{Ewans2000} M. E. Ewans,\ \emph{Braz. J. Phys.} \textbf{30}, 42
(2000).

\bibitem{rdasilva2015} R. da Silva, A. Hentz, A. Alves, \emph{Physica A}, 
\textbf{437}, 139 (2015).

\bibitem{rdasilva2019} R. da Silva, E. V. Stock, \emph{Phys. Rev. E}, 
\textbf{99}, 042148 (2019).

\bibitem{DZJ2018} Ding Z.J., Yu S.L, Zhu K., Ding J.X, Chen B., Shi Q., Lu
X.S, Jiang R., Wang B.H., \emph{Physica A} \textbf{492}, 1700-1714 (2018).

\bibitem{Feller1966} W. Feller, An Introduction to Probability Theory and
Its Applications, New York, J. Wiley (1966).

\bibitem{Gini1921} C. Gini, \emph{Econom. J.} \textbf{31}, 124 (1921).
\end{thebibliography}
\end{document}